# Cracking in polymer glasses and evolution at zero stress: Highlighting discrete long time scale relaxations

S. Mbarek[1,2], P. Baroni[3], L. Noirez[3a)]

[1] *MINES ParisTech, Centre de Mise en Forme des matériaux (CEMEF), 06904 Sophia Antipolis Cedex France*

[2] *Laboratoire de Mécanique de Sousse (LMS), LR11ES36, Ecole Nationale d'Ingénieurs de Sousse, Tunisie*

[3] *Laboratoire Léon Brillouin (CEA-CNRS), CE-Saclay, 91191 Gif-sur-Yvette Cédex France*

Fracture initiation in glassy polymers with no notch is studied together with the evolution at zero stress in the glassy state. Confocal microscopy observations and auto-correlation methods are used to characterize specimens of polymethyl methacrylate (PMMA) loaded at room temperature and subsequently unloaded. The evolution of the morphology and the location of the cracking submitted to elongation rate up to 8% are reported and analyzed during the zero-stress relaxation. The crackling (longitudinal crack, transverse crack and crazes) takes place mainly within a 10µm thickness layer from the surface and does not extend in the bulk. It is shown that the strain field continues to evolve without stress, and that it can be described by an intermittent retraction of the displacement field. Correlatively the number of crazes or of micro-cavities is increasing in the post-loading state at zero-stress relaxation. The timescales involved in the retraction are of the order of several days reporting thus on so far unknown very slow relaxation timescales.

**I. INTRODUCTION:**

From seismic situations [1] down to nanoscale observations [2], the stress action on solid materials is visible via multiple fracture mechanisms due to displacement discontinuity surfaces within the solid. These displacements are visible via longitudinal cracks, dislocations, interfacial sliding, shear cracks, crazes and presents sometimes successive crackling processes called "avalanches" [3]. Localization of these strain and stress zones and the description of their evolution are essential pre-steps to model the deformation of solid materials. In the case of glass polymers, despite more than several decades of studies [4-8] involving complementary techniques as transmission electron microscopy [9,10], electron diffraction, and x-ray scattering [11-13], a continuum model is still to be established. For such models to be successful, the microstructure of the material is the key point requiring in turn further experimental characterization. In this paper, confocal microscopy is used to image deformation zones in millimeter thick samples of polymethylmethacrylate (PMMA) while they deform under uniaxial tensile stress and after loading, at zero stress relaxation. PMMA is classified as a brittle amorphous thermoplastic polymer. The mode of yielding is a dilatational process via the formation of crazes. The craze is a localized plastic zone that develops above a critical stress level, giving rise by cavitation to a local dilatation via void creation of oriented fibrillation [14]. At sufficiently high stress, the fibrils break accelerating the cracking [15-21].

a) Electronic mail : laurence.noirez@cea.fr

In this letter, we mainly focus on the deformation of the glassy polymer arising once the loading is released using confocal microscopy and recent developments in image correlation. A few selected results are presented proving the evolution of the glass at zero-stress relaxation as soon the lowest elongation rates (1%). We highlight that the microfractures continue to nucleate and evolve at zero applied stress. Moreover it is shown that the glassy state relaxes from stresses via discrete relaxation timescales.

## II. EXPERIMENTAL

The material selected is a commercial grade of polymethyl methacrylate (PMMA) provided as sheets of 1 mm thickness and 63 mm length. All the measurements are carried at room temperature in the glassy state; i.e. away from the glass transition temperature ($T_g$ = 100°C). Samples are manufactured in the shape of ISO1/2 test bars in accordance with the NFT 51–034 standard. A home-made stretching device was built for tensile tests and in-situ microscopy observations. The displacement was manually applied. The tension is applied to the specimen though clamps whose symmetry avoids any parasitic out of plane curvature and which are articulated in order to self-align and avoid in plane curvature effects too. The elongation rate $\lambda = (L-L_0)/L_0$ is deduced from the number of the translation device screw rotations and from the sample elongation with respect to its initial length $L_0$. The samples are assumed to be in an equilibrated state prior the experiment and have not been subjected a mechanical pre-treatment. The microscopic observation is carried out with an optical microscope (Olympus BX60) equipped with a Sony CCD camera. The prior observation confirms the absence of defects and cracking. The micrographs have been recorded in transmission mode (Fig.1) and in con-focal geometry. The depth of field (DOF) is ±4µm. AFM measurements were carried in taping mode using a Bruker Dimension Icon Atomic Force Microscope. The image correlation analysis is carried out using the software *DirectPIV* from R&D Vision.

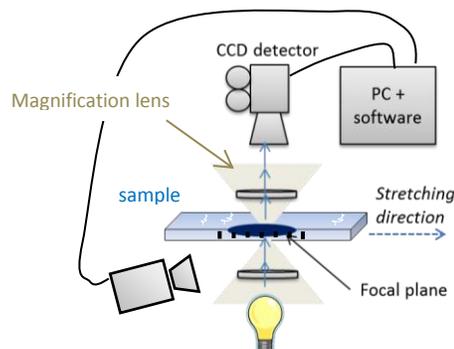

FIG.1: Scheme of the microscopic observation setup in transmission mode.

## III. RESULTS

The mechanical behavior and the related microstructural changes induced in the glassy state of PMMA under uniaxial loading are well referenced in the literature [22].

We remind here the principal deformation zones arising by applying uniaxial stress in the glassy state of amorphous polymers as the PMMA. At low elongation rates, the major population consists in microcavities appearing on the surface. As the level of loading increases, the micro cavities reach a critical amount and some of them transform into crazes developing perpendicular to the stretching axis (Fig.2). These crazes often concentrate along millimeter-length defect lines extended along the stretching axis. These primary defect lines extended along the stress direction favor the initiation of smaller orthogonal cracking including micro-cavities and crazes (Fig.2). The cracking evolve under loading in dimensions and numbers after an induction time when the load is increased in agreement with previous observations [15,22].

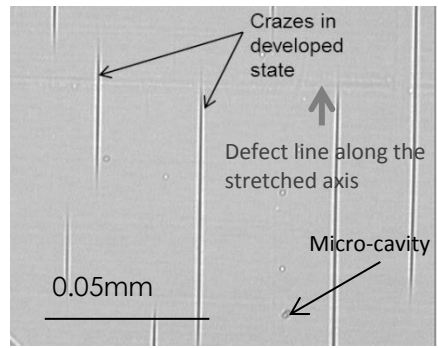

FIG.2: Illustration of typical damages appearing under loading: Micrograph (illuminated from the back with a polychromatic light) of a PMMA specimen submitted to a horizontal uniaxial stretching.

In the present letter, we take advantage of the optical transparency of the sample to operate in transmission mode. This geometry enables the exploration in the sample deepness, to correlate bulk and surfacic deformations and to build a detailed 3D profile of the deformation zone using the scattered light intensity. Combined to an auto-correlation analysis of the time-dependence evolution of the micrographs, we highlight post-loading events that occur at zero-stress in amorphous PMMA indicating that the loading energy can be restored several days after the sample was stretched. This outcome is important given the wide use of these glassy materials.

The exploration of the PMMA sample in its deepness is carried out using the confocal microscopy. It reveals that the damage processes due to the loading takes place mainly within 10μm from the surface (the average sample thickness is 1mm). The surfaces are thus particular deformation zones where the stress releases partially by multiplying the generation of localized defects.

To describe the kinetics of the sample deformation in its bulk and eventually locate a specific surface or interfacial mechanism, the sample is observed from its slice. To facilitate the observation, the sample slice is decorated on its thickness with straight markers (insert of Fig.3) those positions are recorded during the elongation process and analyzed by the technique of image correlation. The aim is to follow the displacement and the deformation of these markers under loading. Fig.3 describes the position of a marker along the stretching axis as a function of time (blue points). Without loading, the marker does not deviate from its initial position. The deviation measures the displacement of the marker from the initial position which is here about 1%. This displacement is in agreement with the applied macroscopic elongation of $\lambda = 1\%$. Since no distortion of the marker is observable during its displacement from the beginning to the end of the loading, we can conclude that the stretching is homogeneously transmitted to the bulk without producing a noticeable layering process due to an interfacial displacement at the surface. This is confirmed by the simultaneous displacement of the markers. In other words, the cracking observable on the surface is directly related to what is happening in the bulk.

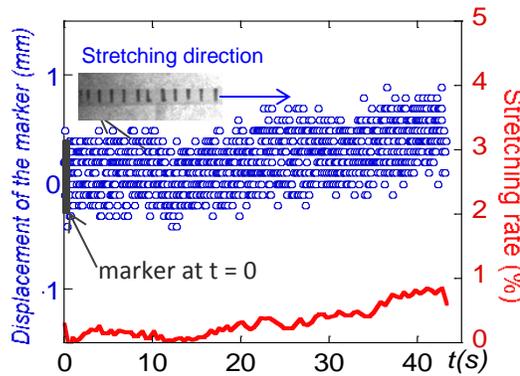

FIG.3: Plot of the displacement of a marker decorating the sample slice (and schemed by a grey bar at $t = 0$) along the stretched axis with respect to its initial position. Data recorded *in situ* during the elongation process. Similar displacements are observed for each marker indicating that the displacement field is homogeneous within the bulk (not illustrated).

We now examine how the cracking process relaxes from residual stresses after loading: the post-loading evolution at the zero stress relaxation. Most of the relaxation studies in glasses consist in a measurement at fixed deformation and are thus generally recorded in a stretched state [23], inducing relaxation processes over a broad spectrum of time scales. A major work from Pruitt et al [24] describes the evolution of the microstructure in polystyrene samples due cyclic stress fields and evidences the role of residual stresses which are interpreted as clamping forces exerted during unloading. These forces generate a zone of residual compressive stresses around the crack. In the present study, we focus on the time-dependent behavior of the PMMA sample at long relaxation times from stress release after stress solicitation in the unloaded state (the zero-stress relaxation state).

Figure 4 shows two snapshots of the PMMA sample recorded in the zero-stress relaxation regime after an elongation rate of $\lambda = 3\%$. The micrographs are recorded immediately and 1 hour later respectively, after having removed the stress. The number of micro-cavities increases during the first hour at a rate of about 80 micro-cavities/mm$^2$/hour at zero-stress relaxation. These localized events demonstrate that the dynamic of damage in glass materials continues after release from the applied stress and develops with time.

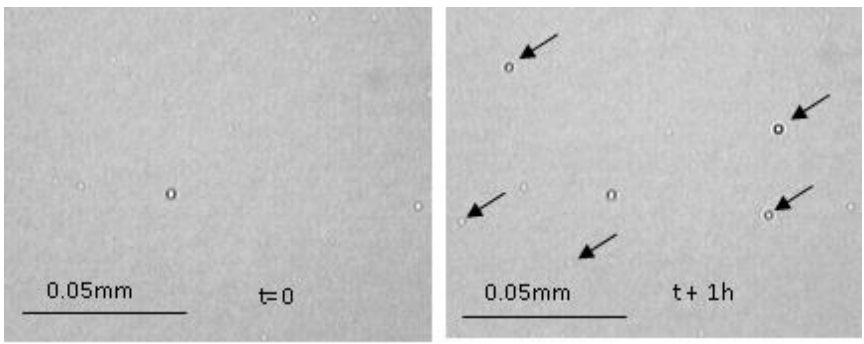

FIG.4: Zero-stress relaxation micrograph recorded: a) immediately after a loading of 3% b) after 1 h. The observations are carried out in the same zone. The stretching was applied along the horizontal direction.

Similarly, the zero-stress micrographs recorded immediately and after 40 hours, after a stretching of 6% (Fig.5) exhibits in addition to micro cavities, some crazes appearing perpendicular to a "phantom" stretched axis. The glass polymer releases the loading energy via a subsequent creation of oriented cracking keeping the memory of the stretched axis.

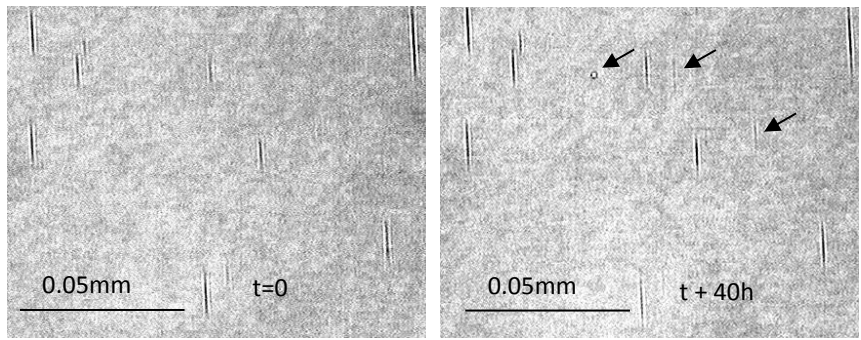

FIG. 5: Zero-stress relaxation micrograph recorded: a) t = 0 s immediately after an elongational rate $\lambda = 6\%$. b) after $t = 40$ hours. The arrows indicate the additional microcavities and crazes. The observations are carried out in the same zone. The stretching was applied along the horizontal direction.

Similar observations are carried out on all the tested samples revealing the generic character of the zero-stress evolution of the crackling. Both the number and the shape of the cracking increase after removing the stress. These "zero-stress" crazes are characterized by a kinetic of several hours. They all develop perpendicularly to the stretched axis similarly as the ones submitted to the elongation stress.

We now examine the morphology of the cracking (Fig.6a). The comparison of micro-snapshots recorded at different times confirms not only the creation of cracking but also a further evolution of the existing cracking. The micrographs are analyzed taking profit from the transmission mode recording. The transmitted intensity is directly related to the mean light path of the light through the sample; the interface created by cracking defects scatters the beam via multiple reflections [25]. An excess of material thus amplifies the transmitted light whereas a lack (hollow) of material absorbs it. This representation enables to evidence easily the excess of material at the borders of the craze and a lack of material in the middle of the craze (Fig.6b). Fig.6c displays a transverse cut of the intensity in the middle of the craze measured on the surface and at 10µm below (confocal geometry). The profile points out a net compensation between the loss and the excess of material and a widening of the hollow under the surface as agreement with previous descriptions [26]. This 3D-optical profile is corroborated by surface AFM profilometry that scows two peaks separated by a hollow (fig.6d). The AFM profilometry is however limited by the cantilever geometry that cannot probe depths larger than 100nm.

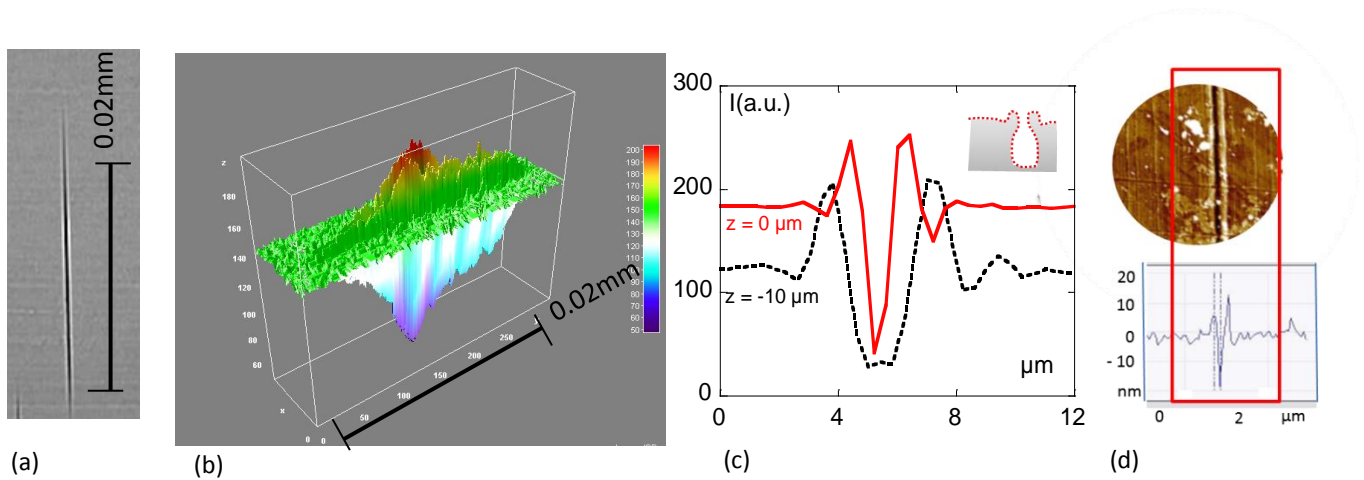

(a) (b) (c) (d)

FIG.6: (a) Typical micrograph (polychromatic light) of a craze. (b) 3D representation of the intensity levels of the transmitted light (ADU units) corresponding to a craze. The stretching was applied along the z direction. (c) Cut of the intensity in the middle of the craze (along the z axis) recorded at the surface (red curve) and at 10µm below (black dotted points). (d) AFM profilometry: 2D mapping of a craze and corresponding transverse profile showing the excess of material at the borders and a lack in the center in agreement with the optical analysis. Similar morphology with a mid-hollow and an excess at the borders is observed both under loading, during the relaxation and for crazes created at zero-stress relaxation.

We use the optical micrographs to compare the 3D optical transmission profiles recorded during the zero-stress relaxation process. Figure 7 illustrates the profile of a cracking immediately after the load is removed and after 4 hours.

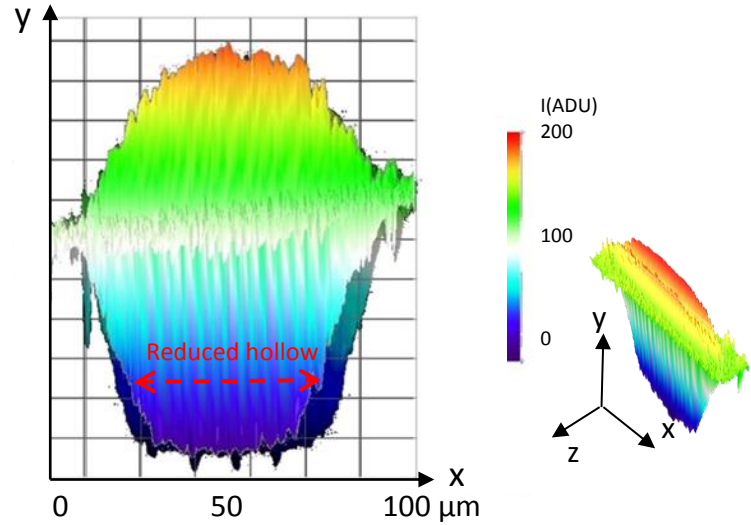

FIG.7: Self-healing process at zero-stress relaxation. Superposition of the side-views of a craze recorded in the zero-stress relaxation regime, immediately after a 10% elongation rate (back profile) and after 4 hours (front profile). The most remarkable change is the width of the hollow part (blue area at the bottom) which reduces in time (here about 18% after 4 hours relaxation). The stretching ($\lambda = 8\%$) was applied along the z axis.

The excess of material at the two borders is in warm colors and the hollow in the middle of the craze corresponds to the blue area. The careful examination of Fig.7 indicates that the width of the hollow part reduces in time. The post-loading evolution at zero-stress corresponds to a self-healing consisting in a partial filling of existing cracking. This local re-arrangement might indicate a retraction of the material which in turn may be at the origin of new cracks.

The retraction process in the post-loading regime at zero-stress can be also pointed out by recording the position of a point chosen on the surface of the sample as a function of the time (at long time scales). If this position does not evolve with time (or evolves randomly), its auto-correlation function is flat. If not, the auto-correlation function measures the displacement from its initial position. This method enables to identify if and how variables recorded at different times are correlated. The spatial correlation function can be expressed by: $\Phi(m,n) = \sum_{i=0}^{M-1} \sum_{j=0}^{N-1} g_1(i,j) g_2(i+m, j+n)$ where $g(i,j)$ is the signal (intensity) measured at the position *(i,j)* of the space, determined by comparing the displacement of the given position *(i,j)* on a microscopic snapshot with respect to its position *(i+m, j+n)* on the next snapshot and so on. From the elementary pixel size and the time delay between two photographs, the auto-correlation function is converted in a displacement field versus time.

Figure 8 shows the displacement field after an elongation rate of 1% and 8% respectively. The direction of the displacement field indicates a retraction of the sample along the former elongation axis. At $\lambda=1\%$ elongation rate, the modeling of the retraction can be expressed by a simple relaxation time and a single length retraction: $\Delta d_{1\%}(t) = - d_{1\%}.(1-exp(t/\tau_{1\%}))$ with $d_{1\%} = 1.5$ hour and $d_{1\%} = -6.78\mu m$. This micron-size retraction corresponds to a tiny quantity about 1% of 1% elongation rate; i.e. 0.01% of the total length retracts. At $\lambda= 8\%$, a simple relaxation modeling is no more possible and the best modeling of the behavior requires at least two relaxation functions: $\Delta d_{8\%}(t) = - d_1.(1-exp(t/\tau_1)) + d_2.(1-exp(t/\tau_2))$ where $d_1$ and $d_2$ are two retraction distances and $\tau_1$ and $\tau_2$ the associated characteristic retraction times. In the present experimental conditions and independently of the location in the sample, the modeling indicates that the retraction distances are rather similar for the two relaxation function $d_1 \cong d_2 \cong -21.5\mu m$, but corresponds to two distinct time scales: $\tau_1=11$ min and $\tau_2 =2.8$ hours, thus indicating that the retraction is at least a two-step process. The observation of discrete relaxation time scales is interesting. It presents analogies with certain systems under stress (as deformed metals, fault seismicity…) governed by avalanche processes, also named crackling dynamics [28]. In the present case, the identification of separated relaxation time scales might indicate that the relaxation can be also governed by slow cracking dynamics. Several measurements located at different places of the sample reproduce a similar relaxation displacement field as illustrated in the insert of Figure 9 for the retraction after a 8% elongation rate.

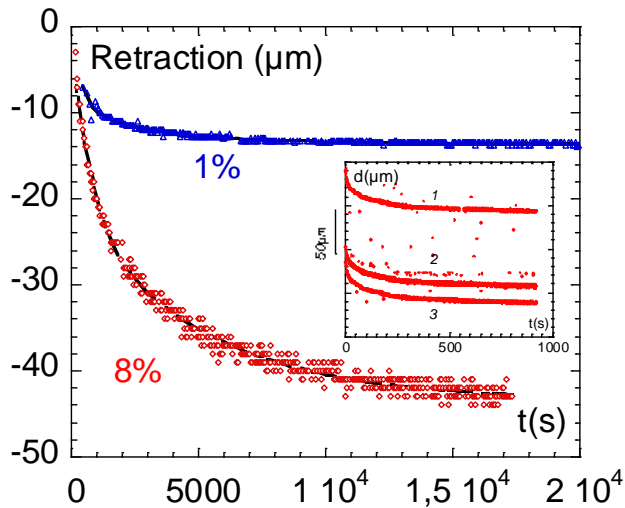

FIG.8: Evolution of the average displacement field at zero stress (retraction) of PMMA samples previously stretched with an elongation rate $\lambda = L/L_0$. Blue triangles: retraction after $\lambda = 1\%$ elongation rate, red lozenges: retraction after $\lambda = 8\%$ elongation rate. The continuous black lines indicate the modeling. The inset displays the retraction on different places of the sample at zero stress after 8% elongation rate.

## IV. CONCLUSION:

Confocal microscopy, AFM and auto-correlation methods have been used to analyze the occurrence of cracking in a polymer glass (PMMA) submitted to an elongation stress and in the relaxation state. The confocal analysis shows that the crazes and micro-cavities mainly develop within a 10µm layer from the surface and are absent in the bulk. The 3D-description of the cracking morphology indicates an excess of material at the borders of the craze and a loss in its center in agreement with the AFM profilometry result.

The observation of the kinetics of sample deformation from its slice indicates that the displacement field is homogeneous within the sample. In the post-loading regime at zero stress relaxation, the fine analysis of the micrographs that the dynamic of damage in glass materials continues to evolve once the applied stress is released by the appearance of new micro-cavities and crazes with time. This kinetic of evolution of the strain field in the relaxation state can be characterized by an auto-correlation function of the displacement field. This displacement field indicates a retraction of the sample. The retraction lengths are weak of the order of several microns. These tiny retraction lengths with respect to the elongation rate plays a crucial role in the glassy solid and are likely responsible of the creation of new cracking. The retraction after a low elongation rate (1%) can be modeled with a single relaxation function while after a larger elongation rate (8%), the displacement field indicates that the retraction of the sample occurs in several steps corresponding to a sum of several relaxation functions defined by well separated relaxation times scaling of several minutes and hours respectively. These relaxation time-scales confirm the existence of heterogeneous dynamics, governing the post-stress behavior and participating to the mechanism of failures. The sample returns back to its equilibrium via an activated hopping process. We anticipate that this experimental observation has profound consequences implying physical and mechanical properties of glass materials: elastic moduli, yield stress and post-yield response, elongation to break threshold, fracture, fatigue, impact strength…

## ACKNOWLEDGMENTS

The authors are very grateful to Bruker for allowing the test of the samples with their upgraded AFM equipment "Dimension Edge Atomic Force Microscope", R&D Vision for the software enabling the computing of the displacement field and to Elisabeth Bouchaud for fruitful discussions.

## REFERENCES

[1] A. Obermann, B. Froment, M. Campillo, E. Larose, T. Planès, B. Valette, J.H. Chen, Q.Y. Liu, J. of Geophyical Research: Solid Earth, 119, 3155, 2014.
[2] Bin Zhang, Lanjv Mei, Haifeng Xiao, App. Phys. Lett. 101, 121915, 2012.
[3] J. Bares, L. Barbier, D. Bonamy, Phys. Rev. Lett. 111, 54301, 2013.
[4] E.J. Kramer and L.L. Berger, Adv. Polym. Sci. 1, 91, 1990.
[5] A.S. Argon and J.G. Hannoosh, Philos. Mag. 36, 1195, 1977.
[6] B.D. Lauterwasser, E.J. Kramer, Philos. Mag. 39, 469, 1979.
[7] E.J. Kramer, Adv. Polym. Sci. 1, 52, 1983.
[9] J. Washiyama, C. Creton, E.J. Kramer, Macromolecules, 25, 4751, 1992.
[10] G.H. Michler, Colloid Polym. Sci. 264, 522, 1986.
[11] E. Paredes, E.W. Fischer, Makromol. Chem, 180, 2707, 1979.
[12] H.R. Brown, E.J. Kramer, J. Macromol. Sci. Phys. B19, 487, 1981.
[13] M.G.A. Tijssen, E. Van der Giessen, L.J. Sluys, Int. J. Solids Struct. 37, 7307, 2000.
[14] L. Wenbo, L. Wenxian, Polymer Testing. 26, 413, 2007.
[15] H.H. Kausch (Ed.), Crazing in polymers, vol. 2, Advances in Polymer Science, vol. 91/92, Springer, Berlin, 1990.
[16] T.N. Krumenkin, G.H. Fredrickson, Macromolecules, 32, 5029, 1999.
[17] W. Luo, T.Q. Yang, X. Wang, Polymer 45, 3519, 2004.
[18] R. Marissen, Polymer 41, 1119, 2000.
[19] I. Narisawa, A.F. Yee, Crazing and Fracture of Polymers, in: E.L. Thomas (Ed.), Structure and properties of polymers, Materials Science and Technology, A comprehensive Treatment, vol. 12, VCH, Weinheim, 1993, 698–765.
[20] R. Schirrer, Damage mechanisms in amorphous glassy polymers: crazing, in: J. Lemaitre (Ed.), Handbook of Materials Behavior Models, Academic Press, San Diego, 2001, pp. 488–500.
[21] S. Socrate, M.C. Boyce, A. Lazzeri, Mech. Mater. 33, 155, 2001.
[22] A. S. Argon, M. M. Salama, Philos. Mag. 36, 1217, 1977.
[23] H.R. Brown, Mater. Sci, Rep. 2, 315, 1987.
[24] C. Creton, E.J. Kramer, H.R. Brown, C.-Y. Hui, Adv. Polym. Sci. 156, 55, 2001.
[25] K. Binder, W. Kob, Glassy Materials and Disordered Solids: an Introduction to their Statistical Mechanics (Singapore: World Scientific), 2005.
[26] K. Thyagarajan, A. K. Ghatak, Fiber Optic Essentials, Wiley-Interscience, 2007.
[27] L. Pruitt, S. Suresh, Polymer, 35, 3221, 1994.
[28] J.P. Sethna, K.A. Dahmen, C.R. Myers, Nature, 410, 242, 2001.